\documentclass[showpacs,]{revtex4}
\usepackage{bm}
\usepackage{amsmath}
\usepackage{epsfig}
\usepackage{graphicx}
\usepackage{color}

\begin{document}

\title{Relativistic calculation of the pion loop correlation energy in nuclear matter in a  theory including confinement}
\author{\'E. Massot, G. Chanfray} 
\affiliation{IPN Lyon, Universit\'e de Lyon, Univ.  Lyon 1, 
 CNRS/IN2P3, UMR5822, F-69622 Villeurbanne Cedex}
\begin{abstract}
We present a relativistic calculation of the saturation properties of nuclear matter which contains the correlation energy. Pion loops are incorporated on top of a relativistic Hartree-Fock (RHF) approach based on a chiral theory. It includes  the effect of nucleon structure through its response to the background chiral invariant scalar field. All the parameters which enter the RHF calculation are fixed or strongly constrained by hadron phenomenology or lattice data. The new input  for the correlation energy is  the Landau-Migdal parameter $g'$ governing the short-range part  of the spin-isospin interaction. We find that the inclusion of the correlation energy improves the description of the saturation properties of nuclear matter.
\end{abstract}

\maketitle
\section{Introduction}\label{Intro}
In a set of recent papers \cite{CE05, CE07, MC08} we have proposed a relativistic description of nuclear matter where the nuclear binding is obtained 
with a chiral invariant background field governing  the medium modifications  of the non-pionic piece of the chiral condensate. As a starting point we take the point of view that the effective theory has to be  formulated in terms of the fields associated with the fluctuations of the chiral quark condensate parametrized in a matrix form, $W=\sigma + i\vec\tau\cdot\vec\pi$. The dynamics of these fields is governed by an  effective potential $V(\sigma, \vec{\pi})$ having a  typical mexican hat shape reflecting  a broken (chiral) symmetry of the QCD vacuum. Such an effective theory can be seen as emerging  from a low (space-like) momentum expansion of a bozonized Nambu-Jona-Lasinio action and the connection  of the scalar "sigma" field $\sigma$ with physical scalar mesons is not necessarily implied  \cite{CHS01}. Explicit application to the description of nuclear matter of such a NJL model has been performed in ref. \cite{BT01}.

As proposed in a previous paper \cite{CEG02} an alternative and very convenient formulation of the resulting sigma model is obtained by going from cartesian to polar coordinates  {\it i.e.}, going from a linear to a non linear representation,  according to~:
$W=\sigma\, + \,i\vec\tau\cdot\vec\pi=S\,U=(f_\pi\,+\,s)\,\exp\left({i\vec\tau\cdot\vec\varphi_\pi/ f_\pi}\right)$.
The new pion field $\vec\varphi_\pi$ corresponds to an orthoradial soft mode which is automatically massless (in the absence of explicit chiral symmetry breaking) since it is associated with rotations on the chiral circle without cost of energy. 
The new sigma meson field $S$, which is a chiral invariant,  describes a radial mode associated with the fluctuations of the ``chiral radius'' around its vacuum expectation value $f_\pi$. This expectation value plays the role of a chiral order parameter around the minimum of the effective potential and  the medium can be seen as a shifted vacuum. With increasing density, its fluctuation $s=S-f_\pi$ governs   the progressive shrinking of the chiral circle and   the evolution of the nucleon mass. Our main physical assumption proposed  in ref. \cite{CEG02} consists in identifying  this chiral invariant $s$ field with the sigma meson of nuclear physics and relativistic theories of the Walecka type, or, said differently, with the background attractive scalar field at the origin of the nuclear binding. This also gives a plausible answer to the long-standing problem of the chiral status of Walecka theories \cite{SW86}.

It is nevertheless well known that in such chiral theories, independently of the details of the modelling, tadpole diagrams  associated with  the mexican hat potential automatically generate attractive three-body forces destroying saturation \cite{KM74,BT01}. 
The origin of this failure can be attributed to the neglect of the effect of nucleon substructure linked to the confinement mechanism as already pointed out in some of our previous works \cite{CE05,CE07,EC07}. It was demonstrated that nuclear matter stability can be restored once the scalar response of the nucleon depending on the quark confinement mechanism is properly incorporated in a way inspired from the QMC model \cite{G88}. 
In particular it was shown that a sufficiently large scalar nucleon response supported by lattice data \cite{LTY04} is able  to generate three-body repulsive forces overcompensating the attractive ones coming from chiral tadpoles \cite{EC07}.  This theoretical framework has been implemented in nuclear matter calculation at the Hartee level \cite{CE05}. In a subsequent work  \cite{CE07} we also incorporated non relativistically the pion loop correlation energy. A full relativistic Hartree-Fock (RHF) calculation was then done in \cite{MC08} allowing to reproduce also the asymmetry properties of nuclear matter. The aim of this paper is to combine these results and to provide a fully relativistic calculation of the pion loop correlation energy on top of the RHF calculation. 

Our article is organized as follows. The second section is a    brief summary of our previous works and gives the main results of the RHF approach. Section 3 is devoted to the formalism that we utilized for the calculation of the polarization propagators entering the correlation energy.  Finally in section 4 we make the connection of the present work with our previous non relativistic calculation of the pion loop energy and  numerical results are given and discussed.

\section{Summary of the RHF approach}
In our previous paper \cite{MC08} we used the following lagrangian which includes the effect of the scalar field $s$, omega, rho and pion exchanges
\begin{equation}
{\cal L}=\bar\Psi\,i\gamma^\mu\partial_\mu\Psi\,+\,
{\cal L}_s\,+\,{\cal L}_\omega\,+\,{\cal L}_\rho\,+\,{\cal L}_\pi
\end{equation}
with
\begin{eqnarray}
{\cal L}_s &=& -M_N(s)\bar\Psi\Psi\,-\,V(s)\,+\,\frac{1}{2} \partial^\mu s\partial_\mu s\nonumber\\	
{\cal L}_\omega &=& - g_\omega\,\omega_\mu\,\bar\Psi\gamma^\mu\Psi\,+\,\frac{1}{2}\,m^2_\omega\,\omega^\mu\omega_\mu
\,-\,\frac{1}{4} \,F^{\mu\nu}F_{\mu\nu}\nonumber\\	
{\cal L}_\rho &=&- g_\rho\,\rho_{a\mu}\,\bar\Psi\gamma^\mu \tau_a\Psi
\,+\,g_\rho\frac{\kappa_\rho}{2\,M_N}\,\partial_\nu \rho_{a\mu}\,\Psi\bar\sigma^{\mu\nu}_{}\tau_a\Psi
\,+\,\frac{1}{2}\,m^2_\rho\,\rho_{a\mu}\rho^{\mu}_{a}
\,-\,\frac{1}{4} \,G_a^{\mu\nu}G_{a\mu\nu}\nonumber\\		
{\cal L}_\pi &=& \frac{g_A}{2\,f_\pi}\,\partial_\mu\varphi_{a\pi}\bar\Psi\gamma^\mu\gamma^5\tau_a\Psi
-\,\frac{1}{2}\,m^2_{\pi}\varphi_{a\pi}^2
\,+\,\frac{1}{2}\, \partial^\mu\varphi_{a\pi}\partial_\mu \varphi_{a\pi}.
\end{eqnarray}
The form of $M_N(s)$ 
\begin{equation}
	M_N(s)=M_N\,+\,g_S\,s\,+\,\frac{1}{2}\,\kappa_{NS}\,\left(s^2\,+\,\frac{s^3}{3\,f_\pi}\right)\label{NUCLEONMASSCORR}
	\end{equation}
reflects the internal nucleon structure through the presence of the scalar response of the nucleon, $\kappa_{NS}$, which can be constrained by lattice data analysis  \cite{CE07}. According to this analysis we obtained an estimate of the dimensionless parameter: $C= \left({f^2_{\pi}/  2\,M_N} \right)
\kappa_{NS}\simeq 1.25 $. This parameter governs the $s^2$ contribution to the
in-medium nucleon mass and generates three-body repulsive forces which are
necessary to get nuclear matter stability and saturation. The scalar mass
parameter $m_\sigma=800$ MeV is also taken from the lattice data analysis. In
addition  the nucleon mass may very well have higher order derivatives with
respect to the scalar field. In practice, as in our previous works
\cite{CE05,CE07}, we also introduce a cubic term. Hence the scalar
susceptibility becomes density dependent~:
\begin{equation}
\tilde\kappa_{NS}(s)={\partial^2M_N\over\partial s^2}=
\kappa_{NS}\left(1\,+\,{s\over  f_\pi}\right)
\end{equation} 
and vanishes at full restoration, $\bar s=-f_\pi$, where $\bar s$ is the expectation value of the $s$ field. 
Hidden in the above Lagrangian is the explicit chiral symmetry breaking piece
\begin{equation}
{\cal L}_{\chi SB}=c\,\sigma=-\frac{c}{2}\,Tr (f_\pi\,+\,s)\,\exp\left({i\vec\tau\cdot\vec\varphi_\pi/ f_\pi}\right)
\simeq \,c\,s\,-\,\frac{c}{2\,f_\pi}\,\varphi_\pi^2	
\end{equation}
which generates the pion mass term with the identification $c=f_\pi\,m_\pi^2$. It is thus implicit that neglecting the higher order terms in the exponent,  the self-interactions of the pions are omitted. Notice that the only meson having a self-interacting potential $V(s)$ is the scalar meson $s$. We take it in practice as in the linear sigma model with the inclusion of the explicit chiral symmetry breaking piece~:
\begin{eqnarray}
	V(s)&=& {\lambda\over 4}\,\big((f_\pi\,+\,s)^2\,-\,v^2\big)^2\,-\,f_\pi m_\pi^2\,s\nonumber\\
	&\equiv &\frac{m^{2}_{\sigma }}{2}\,s ^{2}\,+\,
\frac{m^{2}_{\sigma}\,-\,m^{2}_{\pi }}{2\,f_\pi}\,s^3\,+\,
\frac{m^{2}_{\sigma}\,-\,m^{2}_{\pi }}{8\,f_\pi^2}\,s^4.
\end{eqnarray}
The other parameters ($g_\omega, g_\rho, \kappa_{\rho}, g_A$ and the known meson masses) will be fixed 
as most as possible by hadron phenomenology. In particular  we use the Vector Dominance  picture (VDM) which implies 
 the identification of $\kappa_{\rho}$ with the anomalous part of the isovector magnetic moment of the nucleon, {\it i.e.}, $\kappa_{\rho}=3.7$.  The omega meson should also possess a tensor coupling but, according to VDM the 
corresponding anomalous isoscalar magnetic moment is $\kappa_{\omega}=0.13$. Since it is very small we neglect it here. 

The energy density of symmetric nuclear matter writes~:
\begin{equation}
\epsilon=\epsilon_{kin+Hartree}\,+\,\epsilon_{Fock}
\end{equation}
with
\begin{equation}
\epsilon_{kin+Hartree} = \int \frac{4\,d^3{\bf k}}{(2\pi)^3}\,\left({\bf k}\cdot\frac{{\bf k}^*}{E^*}\,+\,M_N(\bar s)\,\frac{M^*}{E^*}\right)
\,+\,V(\bar s) 
 +\frac{1}{2}\left(\frac{g_\omega}{m_\omega}\right)^2\rho^2.\label{EPSKINHA}
 \end{equation}
The equation of motion for the classical scalar field $\bar s$ is~:
\begin{equation}
	-\nabla^2 \bar s \,+\,V'(\bar s)=-g^*_S\,\rho_S\qquad\hbox{with}\qquad
	g^*_S=\frac{\partial M_N(\bar s)}{\partial \bar s}=g_S\,+\,\kappa_{NS}\,\bar s\,+\,...
\end{equation}
The effective nucleon mass ${M^*}$ and the effective momentum ${\bf k}^*$ come from the Hartree-Fock equations. As discussed in ref. \cite{MC08} 
in the non relativistic limit the orbital wave functions are imposed: these are simply non relativistic plane waves. Consequently if the system is  not too  relativistic, the results of the calculations should not depend very much on the choice of the wave functions. Said differently, the results obtained with  another basis than the fully self-consistent Hartree-Fock (HF) basis  would deviate from the HF results only by tiny relativistic effects as shown in \cite{MC08}. For this reason we introduce the Hartree basis which ignores the Fock term in the nucleon self-energy. In that case   the effective mass ${M^*}$ coincides with $M_N(\bar s)$, the effective momentum ${\bf k}^*$ coincides with the normal momentum ${\bf k}$ and 
the effective nucleon energy becomes $E^*_p(\bar s)=\sqrt{p^2\,+\,M^{2}_N(\bar s)}$ . One advantage will be  the strong simplification of the calculation of the polarization propagator entering the calculation of the pion loop correlation energy.

\smallskip\noindent
The Fock contribution from scalar exchange is given by
\begin{equation}
\epsilon_{Fock}^{(s)} = \frac{g_S^{*2}}{2} \int \frac{d^3{\bf k}}{(2\pi)^3}\frac{d^3{\bf k}'}{(2\pi)^3}\,
\frac{1}{({\bf k}-{\bf k}')^2+ m_\sigma^{*2}}\,
\left(1+\frac{M_N^2(\bar s)}{E^*_{\bf k}\,E^{*}_{{\bf k}'}}-\frac{{\bf k}\cdot{\bf k}^{'}}{E^{*}_{\bf k}\,E^{'*}_{{\bf k}'}}\right)\,N_{\bf k}\,N_{{\bf k}'}
\label{FOCKENERGY}
\end{equation}
where $m^{*2}_{\sigma}=V''(\bar s)\,+\,\tilde{\kappa}_{NS}\,\rho_S$ is the effective scalar mass governing the propagation of the scalar fluctuating field. The other Fock term contibutions (omega, rho and pion) to the energy density are listed in the appendix B of ref. \cite{MC08}.	
\section{Formalism for the correlation energy}
For the calculation of the pion loop correlation energy we generalize to the relativistic case the approach used in \cite{CE07}. The correlation energy associated with pion loops is calculated using the well known charging formula method:
\begin{equation}
E_a^{loop} \equiv V\,\varepsilon_{loop}={3}\,V\,\int_{-\infty}^{+\infty} 
{i\,d\omega\over 2\pi}\int{d{\bf q}\over 
(2\pi)^3}\,\int_0^1{d\lambda\over\lambda}
\,V_{a;\mu\nu}(\omega, {\bf q}; \lambda)\,\left(\Pi_a^{\mu\nu}(\omega, {\bf q}; \lambda)\,-\,\Pi_{0a}^{\mu\nu}(\omega, {\bf q}; \lambda)\right)~.\label{ELOOP}
\end{equation}
$V_{a;\mu\nu}(\omega, {\bf q}; \lambda)$ is the residual interaction in the axial-vector channel wich contains together
with the pion exchange potential the effect of Migdal short-range correlations (Landau-Migdal parameter $g'$)
taken in a covariant form ($q^2=\omega^2-{\bf q}^2$) according to \cite{N01,L03}:
$$V_a^{\mu\nu}(\omega, {\bf q}; \lambda)=\lambda^2\,\left(\frac{g_A}{2\,f_\pi}\right)^2\left(\frac{q^\mu q^\nu}{q^2\,-\,m^{2}_{\pi}}\,-\, g'\,g^{\mu\nu}\right)\,v^2(q).$$
In our calculation we will choose a dipole form factor $v(q)$ with a cutoff $\Lambda=0.98$ GeV, such that 
the resulting  contribution to the free nucleon sigma  term, $\sigma^{(\pi)}_N$, is $\sigma^{(\pi)}_N=21.5$ MeV, in agreement with previous determinations \cite{JTC92,BM92,TGLY04}. In the charging formula all the coupling constants at the interaction vertices are systematically weighted with a $\lambda$ factor. This is the origin of the $\lambda^2$ appearing explicitly in the residual interaction. $\Pi_a^{\mu\nu}(\omega, {\bf q}; \lambda)$ is the full polarization propagator in the axial-vector channel in presence of the interactions weighted by $\lambda$. It is defined according to~:
\begin{equation}
\Pi_a^{\mu\nu}(q)=\frac{1}{V}\int	d(t-t')\,d{\bf r}\,d{\bf r}'\,e^{i\,q \cdot(x-x')}\,
\left\langle 0\right|(-i)\,T\bigg(\left(\bar{\psi}\gamma^\mu\gamma^5\psi\right)(x),\,
\left(\bar{\psi}\gamma^\nu\gamma^5\psi\right)(x')\bigg)\left|0\right\rangle .
\end{equation}
Notice that we are actually dealing with correlators in an isovector channel. We omit for simplicity the isospin Pauli matrices  $\tau_j$ since the only effect in symmetric nuclear matter is to multiply at the very end the result for the correlation energy by a factor three. To get  the genuine correlation energy, the first order term, {\it i.e.}, the mean field one, has to be substracted since it is already incorporated as the Fock pion exchange term in presence of short range correlations \cite{MC08}. The explicit expression for these mean-field $\Pi^0$ polarization propagators will be given below.

In the non relativistic limit this correlation energy involves polarization
propagators (or interactions) only in the spin-isospin channel: the pion
exchange interaction is of pure longitudinal spin-isospin nature whereas the
short range interaction contains both  longitudinal and transverse pieces
\cite{CE07}. However, non relativistically, the transverse spin-isospin
polarization propagator is also affected by rho meson exchange. Hence to keep this connection with the non relativistic case \cite{CE07}, we also incorporate the rho meson exchange which is already present in our previous RHF calculation. The rho exchange contributes to the correlation energy according to:
\begin{equation}
E_\rho^{loop} \equiv V\,\varepsilon_{loop}={3}\,V\,\int_{-\infty}^{+\infty} 
{i\,d\omega\over 2\pi}\int{d{\bf q}\over 
(2\pi)^3}\,\int_0^1{d\lambda\over\lambda}
\,V_{\rho;\mu\nu}(\omega, {\bf q}; \lambda)\,\left(\Pi_\rho^{\mu\nu}(\omega, {\bf q}; \lambda)\,-\,\Pi_{0\rho}^{\mu\nu}(\omega, {\bf q}; \lambda)\right).\label{ELOOPRHO}
\end{equation}
$V_{\rho\mu\nu}(\omega, {\bf q}; \lambda)$ is the rho meson exchange interaction:
$$V_\rho^{\mu\nu}(\omega, {\bf q}; \lambda)=-\,\lambda^2\,g^{2}_{\rho}\frac{g^{\nu\nu}}{q^2\,-\,m^{2}_{\rho}}\,v^2(q).$$
For the vector coupling constant we take as in our previous paper the VDM value:
$g_\rho=2.65$. The rho meson propagator should also contain a $q^{\mu}q^{\nu}$
term but it can be omitted since current conservation implies
$q_{\mu}\Pi_\rho^{\mu\nu}=0$. Notice that we take for simplicity the same form factor, {\it i.e}., the same momentum cutoff, as in the pion exhange. We will discuss this point at the end of the paper and show that taking a harder form factor does not significantly alter the conclusions. The vector-isovector polarization propagator is defined (omitting again the isospin Pauli matrices) according to:
\begin{equation}
\Pi_\rho^{\mu\nu}(q)=\frac{1}{V}\int	d(t-t')\,d{\bf r}\,d{\bf r}'\,e^{i\,q \cdot(x-x')}\,
\left\langle 0\right|(-i)\,T\bigg(\left(\bar{\psi}\Gamma^\mu(q)\psi\right)(x),\,
\left(\bar{\psi}\Gamma^{\dagger\nu}(q)\psi\right)(x')\bigg)\left|0\right\rangle .
\end{equation}
The $\rho NN$ vertices are given by:
$$\Gamma_\rho^{\mu}(q)=\gamma^\mu\,-\,i\frac{\kappa_\rho}{2M_N}\sigma^{\mu\alpha}q_\alpha,\qquad
\Gamma_\rho^{\dagger\nu}(q)=\gamma^\nu\,+\,i\frac{\kappa_\rho}{2M_N}\sigma^{\nu\beta}q_\beta .$$
For the tensor coupling we again take the VDM value, $\kappa_\rho=3.7$.
As previously discussed  the axial and vector meson correlators mix (in the non relativistic limit they both contain a transverse spin-isospin piece). Furthermore  we also have to incorporate the mixed polarization propagators~:
\begin{eqnarray}
\Pi_{\rho a}^{\mu\nu}(q)&=&\frac{1}{V}\int	d(t-t')\,d{\bf r}\,d{\bf r}'\,e^{i\,q \cdot(x-x')}\,
\left\langle 0\right|(-i)\,T\bigg(\left(\bar{\psi}\Gamma^\mu(q)\psi\right)(x),\,
\left(\bar{\psi}\gamma^\nu\gamma^5\psi\right)(x')\bigg)\left|0\right\rangle\\
\Pi_{a \rho}^{\mu\nu}(q)&=&\frac{1}{V}\int	d(t-t')\,d{\bf r}\,d{\bf r}'\,e^{i\,q \cdot(x-x')}\,
\left\langle 0\right|(-i)\,T\bigg(\left(\bar{\psi}\gamma^\mu\gamma^5\psi\right)(x),\,
\left(\bar{\psi}\Gamma^{\dagger\nu}(q)\psi\right)(x')\bigg)\left|0\right\rangle .
\end{eqnarray}
The polarization propagators are calculated using a RPA scheme. They are solution of the coupled Dyson equations :
\begin{eqnarray}
\Pi_a^{\mu\nu}(q)&=&	\Pi_{0a}^{\mu\nu}(q)\,+\,\Pi_{0a}^{\mu\alpha}(q)V_{a;\alpha\beta}(q)\Pi_a^{\beta\nu}(q)\,+\,
\Pi_{0a\rho}^{\mu\alpha}(q)V_{\rho;\alpha\beta}(q)\Pi_{\rho a}^{\beta\nu}(q)\nonumber\\
\Pi_{\rho a}^{\mu\nu}(q)&=&	\Pi_{0\rho a}^{\mu\nu}(q)\,+\,\Pi_{0\rho a}^{\mu\alpha}(q)V_{a;\alpha\beta}(q)\Pi_a^{\beta\nu}(q)\,+\,
\Pi_{0 \rho}^{\mu\alpha}(q)V_{\rho;\alpha\beta}(q)\Pi_{\rho a}^{\beta\nu}(q)\nonumber\\
\Pi_\rho^{\mu\nu}(q)&=&	\Pi_{0\rho}^{\mu\nu}(q)\,+\,\Pi_{0\rho}^{\mu\alpha}(q)V_{\rho;\alpha\beta}(q)\Pi_\rho^{\beta\nu}(q)\,+\,
\Pi_{0\rho a}^{\mu\alpha}(q)V_{a;\alpha\beta}(q)\Pi_{a\rho}^{\beta\nu}(q)\nonumber\\
\Pi_{a\rho }^{\mu\nu}(q)&=&	\Pi_{0 a\rho }^{\mu\nu}(q)\,+\,\Pi_{0 a\rho}^{\mu\alpha}(q)V_{\rho;\alpha\beta}(q)\Pi_\rho^{\beta\nu}(q)\,+\,
\Pi_{0 a}^{\mu\alpha}(q)V_{a;\alpha\beta}(q)\Pi_{a\rho }^{\beta\nu}(q).
\label{RPA}
\end{eqnarray}
where the $\lambda$ dependence in the arguments have been omitted for simplicity.
The mean-field polarization propagators entering the above Dyson equations are:
\begin{eqnarray}
\Pi_{0a}^{\mu\nu}(q)&=&-2\int\frac{i d^4p}{(2\pi)^4}
tr_D\left[\,G(p)\,\gamma^\mu\gamma^5\,G(p+q)\,\gamma^\nu\gamma^5\right]\\
  \Pi_{0\rho a}^{\mu\nu}(q)
  &=& 
  -2\int\frac{i d^4p}{(2\pi)^4}
  tr_D\left[\,G(p)\,\Gamma_\rho^\mu \,G(p+q)\,\gamma^\nu\gamma^5\right]
  \\
  \Pi_{0a\rho }^{\mu\nu}(q)
  &=&
  -2\int\frac{i d^4p}{(2\pi)^4}
  tr_D\left[\,G(p)\,\gamma^\mu\gamma^5\, G(p+q)\,\Gamma_\rho^{\dagger\nu}\right]
  \\
  \Pi_{0\rho}^{\mu\nu}(q)
  &=&
  -2\int\frac{i d^4p}{(2\pi)^4}
  tr_D\left[\,G(p)\,\Gamma_\rho^\mu\, G(p+q)\,\Gamma_\rho^{\dagger\nu}\right].
  \label{BARE}
\end{eqnarray}
The mean-field nucleon propagator has the form :
\begin{equation}
G(p)=\frac{1}{p^2 - M^{*2} + i\eta}	+ 2i\pi N_{\bf p} \delta\left(p^2 - M^{*2}\right)\Theta(p^0)
\label{Nprop}
\end{equation}
where $N_{\bf p}$ is the occupation number and it is understood that  the pure vacuum piece of the mean-field polarization propagators is dropped.

To solve the RPA problem and to get a compact form for the correlation energy, we will use a projector technics \cite{CPS92,L03}. For this purpose we introduce the following four-vectors and tensors~:
\begin{equation}
\eta^\mu=(1,{\bf 0}),\qquad \hat{\eta}^\mu= \eta^\mu- \frac{\eta\cdot q}{q^2}q^\mu,\qquad
\hat{\eta}^2=-\frac{{\bf q}2}{q^2}
\end{equation}
\begin{equation}
L^{\mu\nu}=\frac{q^\mu q\nu}{q^2},\quad R^{\mu\nu}=\frac{\hat{\eta}^\mu
\hat{\eta}^\nu}{\hat{\eta}^2},\quad T^{\mu\nu}=g^{\mu\nu}\,-\,L^{\mu\nu}\,-\,R^{\mu\nu}.
\end{equation}
The tensors $L$, $R$ and $T$ satisfy projector properties and  are mutually orthognal. In addition they also satisfy the normalization conditions,  $L^{\mu\nu}L_{\mu\nu}=R^{\mu\nu}R_{\mu\nu}=1$ and $T^{\mu\nu}T_{\mu\nu}=2$. The residual interaction can be decomposed on these covariant tensors:
$$
V_a^{\mu\nu}(q)=\lambda^2\,\left(\frac{g_A}{2\,f_\pi}\right)^2\,v^2(q)\, \bigg[
\left(q^2 D_\pi(q) \,-\,g'\right)\,L^{\mu\nu}\,- \,g' \,T^{\mu\nu}\,-\,	g'\, R^{\mu\nu}\bigg]
$$
\begin{equation}
\equiv  V_{aL}\,L^{\mu\nu}\,+\,V_{aT}\,T^{\mu\nu}\,+\,V_{aR}\,R^{\mu\nu}
\end{equation}
\begin{equation}
V_\rho^{\mu\nu}(q) = - \lambda^2\,g^{2}_{\rho}\,v^2(q)\,\frac{1}{q^2\,-\,m_\rho^2}\,\left(L^{\mu\nu}\,+\,T^{\mu\nu}\,+\,R^{\mu\nu}\right)\nonumber
 \equiv  V_\rho \,\left(L^{\mu\nu}\,+\,T^{\mu\nu}\,+\,R^{\mu\nu}\right).
  \label{}
\end{equation}
The polarization propagators entering the expression of the correlation energy are actually the various projections on the above tensors, namely:
\begin{eqnarray}
& &\Pi_{aL}(q)=L_{\mu\nu}\,\Pi^{\mu\nu}_{a}(q),\qquad \Pi_{aT}(q)=\frac{1}{2}T_{\mu\nu}\,\Pi^{\mu\nu}_{a}(q),\qquad
\Pi_{aR}(q)=R_{\mu\nu}\,\Pi^{\mu\nu}_{a}(q),\nonumber\\
& &\Pi_{\rho L}(q)=L_{\mu\nu}\,\Pi^{\mu\nu}_{\rho}(q),\qquad \Pi_{\rho T}(q)=\frac{1}{2}T_{\mu\nu}\,\Pi^{\mu\nu}_{\rho}(q),\qquad
\Pi_{\rho R}(q)=R_{\mu\nu}\,\Pi^{\mu\nu}_{\rho}(q) .
\end{eqnarray}
The mean-field polarization bubbles have similar projections which means that they can be decomposed according to:
\begin{eqnarray}
& & \Pi^{\mu\nu}_{0a}(q)=L^{\mu\nu}\,\Pi_{0aL}(q)\,+\,T^{\mu\nu}\,\Pi_{0aL}(q)\,+\,R^{\mu\nu}\,\Pi_{0aR}(q)\nonumber\\
& & \Pi^{\mu\nu}_{0\rho}(q)=T^{\mu\nu}\,\Pi_{0\rho L}(q)\,+\,R^{\mu\nu}\,\Pi_{0\rho R}(q).
\end{eqnarray}
We see on these expressions that the rho channel polarization propagator has no projection on the longitudinal channel which  remains true for the full propagator. There is an additionnal mean-field bubble which mixes the axial and rho channels. It has the following tensorial structure:
\begin{eqnarray}
& & \Pi_{0\rho a}^{\mu\nu}(q)=\Pi_{0 a \rho}^{\mu\nu}(q)=T_6^{\mu\nu}\, \Pi_{06}(q)\qquad\hbox{with}\qquad T_6^{\mu\nu}=\frac{i}{|{\bf q}|}
\varepsilon^{\mu\nu\rho\sigma}\hat{\eta}_\rho q_\sigma\nonumber\\
& & T_{6\mu\nu}T^{6\nu\mu}=2,\qquad T_6^{\mu\alpha}T_{6\alpha\nu}=T^{\mu}_{\,\,\nu},\qquad 
T_6^{\mu\alpha}T_{\alpha}^{\,\,\nu}=T_6^{\mu\nu}.
\end{eqnarray}
The antisymmetric $T_6$ tensor is actually orthogonal to the $L$ and $R$ tensors. Consequently the mixing beween the axial and rho channel proceeds only through the transverse channel associated with the tensor $T$. We also introduce the full axial-rho polarization propagators projected on  the $T_6$ tensor
\begin{equation}
\Pi_{\rho a}(q)=\frac{1}{2}T_6^{\mu\nu}\, \Pi_{\rho a; \mu\nu}(q)\qquad
\Pi_{a \rho}(q)=\frac{1}{2}T_6^{\mu\nu}\, \Pi_{a\rho; \mu\nu}(q).
\end{equation}
The correlation energy can be expressed in terms of the projected polarization propagators:
\begin{eqnarray}
E^{loop}&=& E^{loop}_{aL}\,+\,E^{loop}_{T}\,+\,E^{loop}_{aR}\,+\,E^{loop}_{\rho R}\nonumber\\
E_{aL}^{loop}&=&{3}\,V\,\int_{-\infty}^{+\infty}{i\,d\omega\over 2\pi}\int{d{\bf q}\over (2\pi)^3}\,\int_0^1{d\lambda\over\lambda}
\,V_{aL}(\omega, {\bf q}; \lambda)\,\left(\Pi_{aL}(\omega, {\bf q}; \lambda)\,-\,\Pi_{0aL}(\omega, {\bf q}; \lambda)\right)\nonumber\\
E_{T}^{loop}&=&{3}\,V\,\int_{-\infty}^{+\infty}{i\,d\omega\over 2\pi}\int{d{\bf q}\over (2\pi)^3}\,\int_0^1{d\lambda\over\lambda}
\,2\,V_{aT}(\omega, {\bf q}; \lambda)\,\left(\Pi_{aT}(\omega, {\bf q}; \lambda)\,-\,\Pi_{0aT}(\omega, {\bf q}; \lambda)\right)\nonumber\\
&+&{3}\,V\,\int_{-\infty}^{+\infty}{i\,d\omega\over 2\pi}\int{d{\bf q}\over (2\pi)^3}\,\int_0^1{d\lambda\over\lambda}
\,2\,V_{\rho T}(\omega, {\bf q}; \lambda)\,\left(\Pi_{\rho T}(\omega, {\bf q}; \lambda)\,-\,\Pi_{0\rho T}(\omega, {\bf q}; \lambda)\right)\nonumber\\
E_{aR}^{loop}&=&{3}\,V\,\int_{-\infty}^{+\infty}{i\,d\omega\over 2\pi}\int{d{\bf q}\over (2\pi)^3}\,\int_0^1{d\lambda\over\lambda}
\,V_{aR}(\omega, {\bf q}; \lambda)\,\left(\Pi_{aR}(\omega, {\bf q}; \lambda)\,-\,\Pi_{0aR}(\omega, {\bf q}; \lambda)\right)\nonumber\\
E_{\rho R}^{loop}&=&{3}\,V\,\int_{-\infty}^{+\infty}{i\,d\omega\over 2\pi}\int{d{\bf q}\over (2\pi)^3}\,\int_0^1{d\lambda\over\lambda}
\,V_{\rho R}(\omega, {\bf q}; \lambda)\,\left(\Pi_{\rho R}(\omega, {\bf q}; \lambda)\,-\,\Pi_{0\rho R}(\omega, {\bf q}; \lambda)\right).
\end{eqnarray}
The Dyson equations for the projected polarization propagators can be obtained from the original ones (eq. \ref{RPA}) by projecting them on the various channels L, T or R. After straightforward algebraic manipulations one obtains:
\begin{eqnarray}
\Pi_{a L}&=& \Pi_{0a L}\,+\,\Pi_{0aL}\,V_{aL}\,\Pi_{a L}\nonumber\\
\Pi_{a R}&=& \Pi_{0a R}\,+\,\Pi_{0aR}\,V_{aR}\,\Pi_{a R}\nonumber\\
\Pi_{a T}&=& \Pi_{0a T}\,+\,\Pi_{0aT}\,V_{aT}\,\Pi_{a T}\, +\, \Pi_{06}\,V_{\rho}\,\Pi_{\rho aT}\nonumber\\
\Pi_{\rho aT}&=&\Pi_{06}\,+\,\Pi_{ 06}\,V_{aT}\,\Pi_{a T}\, + \,\Pi_{0\rho  T}\,V_{\rho }\,\Pi_{\rho aT}\nonumber\\
\Pi_{\rho T}&=& \Pi_{0\rho T}\,+\,\Pi_{\rho }\,V_{\rho }\,\Pi_{\rho T}\, +\, \Pi_{06}\,V_{aT}\,\Pi_{a\rho T}\nonumber\\
\Pi_{a\rho T}&=&\Pi_{06}\,+\,\Pi_{06}\,V_{\rho}\,\Pi_{\rho T}\, + \,\Pi_{0a T}\,V_{aT}\,\Pi_{a\rho T}\nonumber\\
\Pi_{\rho R}&=& \Pi_{0\rho R}\,+\,\Pi_{0\rho R}\,V_{\rho}\,\Pi_{a R}.
\end{eqnarray}
From the solution we get the particular combinations entering the expression of the correlation energy:
\begin{eqnarray}
V_{aL}\,\Pi_{a L}&=& \frac{V_{aL}\,\Pi_{0a L}}{1\,-\,V_{aL}\,\Pi_{0a L}}\nonumber\\
V_{aR}\,\Pi_{a R}&=& \frac{V_{aR}\,\Pi_{0a R}}{1\,-\,V_{aR}\,\Pi_{0a R}}\nonumber\\
V_{aT}\,\Pi_{a T}\,+\,V_{\rho }\Pi_{\rho T}&=&
\frac{V_{aT}\,\Pi_{0a T}\,+\,V_{\rho }\Pi_{0\rho T}\,+2\,V_{aT}\,V_{\rho }\left(\Pi_{06}^2\, -\,\Pi_{0a T }\,\Pi_{0\rho T} \right)}
{\left(1\,-\,V_{aT}\,\Pi_{0a T}\right)\left(1\,-\,V_{\rho}\,\Pi_{0\rho T}\right)\,-\,V_{aT}\,V_{\rho }\,\Pi_{06}^2}\nonumber\\
V_{ \rho}\,\Pi_{\rho R}&=& \frac{V_{\rho}\,\Pi_{0\rho}}{1\,-\,V_{\rho}\,\Pi_{0\rho R}}.\label{VPI}
\end{eqnarray}
The integration over the varying coupling constant $\lambda$ is immediate and we get the final expression for the various contributions to the 
correlation energy:
\begin{eqnarray}
E^{loop}&=& E^{loop}_{aL}\,+\,E^{loop}_{T}\,+\,E^{loop}_{aR}\,+\,E^{loop}_{\rho R}\nonumber\\
E_{aL}^{loop}&=&- \frac{3 V}{2}\,\int_{-\infty}^{+\infty}{i\,d\omega\over 2\pi}\int{d{\bf q}\over (2\pi)^3}\,
\bigg[\ln\left(1\,-\,V_{aL}\,\Pi_{0a L}\right)\,+\,V_{aL}\,\Pi_{0a L}\bigg](\omega, {\bf q})\nonumber\\
E_{T}^{loop}&=&- 3 V\,\int_{-\infty}^{+\infty}{i\,d\omega\over 2\pi}\int{d{\bf q}\over (2\pi)^3}\,
\bigg[\ln\left[(1-\Pi_{0\rho T }V_{\rho })\,(1-\Pi_{0aT}V_{aT})-V_{aT}\Pi_{06} V_{\rho}\Pi_{06}\right]
 +\Pi_{0\rho T}V_{\rho}+\Pi_{0aT}V_{aT}\bigg](\omega, {\bf q})\nonumber\\    
E_{aR}^{loop}&=&-\,\frac{3 V}{2}\,\int_{-\infty}^{+\infty}{i\,d\omega\over (2\pi}\int{d{\bf q}\over (2\pi)^3}\,
\bigg[\ln\left(1\,-\,V_{aR}\,\Pi_{0a R}\right)\,+\,V_{aR}\,\Pi_{0a R}\bigg](\omega, {\bf q})\nonumber\\ 
 E_{\rho R}^{loop}&=&-\,\frac{3 V}{2}\,\int_{-\infty}^{+\infty}{i\,d\omega\over 2\pi}\int{d{\bf q}\over (2\pi)^3}\,
\bigg[\ln\left(1\,-\,V_{\rho R}\,\Pi_{0\rho R}\right)\,+\,V_{\rho R}\,\Pi_{0\rho R}\bigg](\omega, {\bf q}).  
\end{eqnarray}
The leading term is obtained by expanding the log to  second order in $V\Pi_0$ which yields  the result of second order perturbation theory which is always negative according to a basic result of quantum mechanics. In particular the second order pion loop is embedded in  the axial-longitudinal contribution to the correlation energy. It can be directly compared with the iterated pion exchange (the so-called planar diagram) appearing in medium chiral perturbation theory. If the ChiPT calculation is regularized with a cutoff \cite{KFW02}, a negative result is also obtained but much larger than in our approach where the pion exchange is strongly screened by short-range correlations.

The analytical structure of the polarization propagators considered as a fonction of $\omega$ is such that it has a cut on the real axis and these propagators are analytic in the first and third quadrants. In other words the continuous set of poles lies below the real axis for $\omega$ positive and above the real axis for $\omega$ negative, the same as for the pion and rho propagators. For this reason the (practical) calculation can be done 
using a Wick rotation. Each four momentum integral will be calculated according to:
\begin{equation}
\int_{-\infty}^{+\infty} {i\,d\omega\over (2\pi)}\int{d{\bf q}\over (2\pi)^3}\,	F(\omega, {\bf q})\quad\to\quad
-\int_{-\infty}^{+\infty} {dz\over (2\pi)}\int{d{\bf q}\over (2\pi)^3}\,	F(\omega=iz, {\bf q}).
\label{eqn:WICKrot}
\end{equation}
In practice as an input of the calculation all what we need is the real part of the bare polarization propagators to make an analytical continuation 
to $\omega=iz $. The explicit form of these polarization propagators are given in the appendix.
\section{Discussion and numerical Results}
\paragraph{First order polarization propagators and non relativistic limit.}

The various mean-field projected propagators can be written generically as:
\begin{eqnarray}
\Pi_{0j}(q)&=&- \int{4\,d{\bf p}\over (2\pi)^3}\,\left(\frac{N_{p}}{E_{p}}\,\frac{K_j(q;{\bf p})}{q^2\,+\,2\,p\cdot q\,+\,i\eta}\quad +\quad(q\to -q)\right)
\nonumber\\
& -& \int{4\,d{\bf p}\over (2\pi)^3}\,\,2 i \pi\,\frac{N_p\,N_{{\bf p}+{\bf q}}}{E_p}\,K_j(q;{\bf p})\,\delta\left(q^2\,+\,2\,p\cdot q\right)\Theta(E_p\,+\,\omega)\label{PIFORM}
\end{eqnarray}
with
\begin{eqnarray}
K_{0aL}(q;{\bf p})&=& 2\,M^{*2}\nonumber\\
K_{0aT}(q;{\bf p})&=& 2\,M^{*2}\,+\,p\cdot q\,+\,{\bf p}^2_T\nonumber\\
K_{0aR}(q;{\bf p})&=& -2\,{\bf p}^2_T\nonumber\\
K_{06}(q;{\bf p})&=& \frac{1\,+\,\kappa^*_\rho}{2\, M^*}\,\,2\, M^*\left(\left|{\bf q}\right|\,E_p\,-\,\omega\,{\bf p}\cdot\hat{\bf q}\right)\nonumber\\
K_{0\rho T}(q;{\bf p})&=&-\left(\frac{1\,+\,\kappa^*_\rho}{2\, M^*}\right)^2\,q^2\,2\,M^{*2}\,+\,p\cdot q\,+\,\frac{q^2}{2}\,+\,
{\bf p}^2_T\,\left(1- \left(\frac{\kappa^*_\rho}{2\, M^*}\right)^2\,q^2\right)\nonumber\\
K_{0\rho R}(q;{\bf p})&=& - 2\,M^{*2}\,+\,\left(\frac{\kappa^*_\rho}{2\, M^*}\right)^2\,q^2\,\left(p\cdot q\,+\,2
{\bf p}^2_T\right)\,-\,2\,{\bf p}^2_T\,+\,\kappa^*_\rho\,q^2.\label{KFUNC}
\end{eqnarray}
where $\kappa^{*}_\rho=\kappa_\rho (M^*/M_N)$ appears as  an in-medium modified rho tensor coupling constant and ${\bf p}_T=
{\bf p}-{\bf p}\cdot\hat{\bf q}\,
\hat{\bf q}$, with ${\bf p}^2_T={\bf p}^2\,-\,\left({\bf p}\cdot\hat{\bf q}\right)^2$, is the transverse component of the nucleon momentum.

The difference between relativistic and non relativistic calculations has essentially three distinct origins: the first is the incorporation of nucleon-antinucleon excitations, the second originates from the use of Dirac spinors which translates into the structure of the $K_j$ functions
and the third is pure kinematics (for instance $E_p$ replaced by $M^*\,+\,p^2/2
M^*$). To appreciate the role of  the $\bar N N$ excitations we rewrite the
denominator appearing in eq. \ref{PIFORM} as:
\begin{equation}
\frac{K_j(q)}{E_{p}}\,\frac{1}{q^2\,+\,2\,p\cdot q\,+\,i\eta}=	\frac{K_j(q)}{2\,E_p\,E_{{\bf p}+{\bf q}}}\,
\left(\frac{1}{\omega\,-\,E_{{\bf p}+{\bf q}}\,+\,E_p\,+\,i\eta}\,-\,\frac{1}{\omega\,+\,E_{{\bf p}+{\bf q}}\,+\,E_p\,-\,i\eta}\right).
\end{equation}
 The first term represents the forward going $p-h$ bubble and the second term represents the backward going $\bar N N$ bubble. For the energy denominator  associated with $q$ changed in $-q$, we write similarly:
\begin{equation}
\frac{K_j(-q)}{E_{p}}\,\frac{1}{q^2\,-\,2\,p\cdot q\,+\,i\eta}=	-\,\frac{K_j(-q)}{2\,E_p\,E_{{\bf p}-{\bf q}}}\,
\left(\frac{1}{\omega\,+\,E_{{\bf p}-{\bf q}}\,-\,E_p\,-\,i\eta}\,-\,\frac{1}{\omega\,-\,E_{{\bf p}-{\bf q}}\,-\,E_p\,+\,i\eta}\right).
\end{equation}
The first term represents the backward  going $p-h$ bubble and the second term represents the forward going $\bar N N$ bubble. The inclusion of antinucleons is at best questionable since there are many excitations far below the $\bar N N$ threshold which should be incorporated before. In addition the internal consistency would require to take into account vacuum polarization effects but the restructuring of the QCD vacuum has probably little to do with virtual $\bar N N$ excitations. However at the formal level they are essential to maintain the covariance of the calculation and the simple tensorial structure and symmetry properties ($q_\mu\Pi^{\mu\nu}_\rho =0$). For that reason  we decided to keep them in our calculation as many authors \cite{CPS92,CL99}. Moreover their practical effect is a  small ${\cal O}(1/M_N)$ relativistic correction.

We now come to the various $K_j$ functions (eq. \ref{KFUNC}). The axial longitudinal bubble $\Pi_{0aL}$  (pion channel) is actually combined with $V_{aL}$ which contains the pion exchange whith  an explicit $-q^2={\bf q}^2-\omega^2$ factor. This is a potentially  important difference with the non relativistic approach where the pion coupling contains only ${\bf q}^2$ (pure p-wave coupling). However since the dominant energy range is $\omega\sim {\bf q}^2/2M^*$ (quasi-elastic peak)  the relativistic correction induced at the level of the vertex 
is of order ${\cal O}(1/M_N^2)$. The conventionnal non relativistic limit is obtained by ignoring the antinucleon terms and by replacing 
$1/2E_p\,E_{{\bf p}+{\bf q}}$ by $1/2 M^{*2}$ so that $ {K_j(q)}/E_p\,E_{{\bf p}+{\bf q}}\sim 1$ with the result :
\begin{equation}
\Pi^{NR}_{0aL}=-\Pi_0(\omega, {\bf q}) \equiv	 -\int{4\,d{\bf p}\over (2\pi)^3}\,N_p\,P_{{\bf p}+{\bf q}}\,
\left(\frac{1}{\omega\,-\,E_{{\bf p}+{\bf q}}\,+\,E_p\,+\,i\eta}
-\,\frac{1}{\omega\,+\,E_{{\bf p}+{\bf q}}\,-\,E_p\,-\,i\eta}\right).
\end{equation}
We also see that the $K$ functions for the transverse axial channel and the rho-R channel are governed by the dominant non relativistic term ($\pm 2\,M^{*2}$), the terms in ${\bf p}^2_T$ and $q^2$ being  of ${\cal O}(1/M_N^2)$ order. The term involving $p\cdot q=E_p\,\omega-{\bf p}\cdot{\bf q}$ can also be seen of second order since $\omega\sim {\bf q}^2/2M^*$. The analysis of the rho transverse channel is a little particular since the first term is proportionnal to $q^2$ and apparently of order ${\cal O}(1/M_N^2)$. However the factor $1\,+\,\kappa^*_\rho$ is about $5$. Hence  for typical momentum $\left|{\bf q}\right|\sim 300\,MeV$ the factor 
$(1\,+\,\kappa^*_\rho)/2\, M^*\left|{\bf q}\right|$ turns out to be of order unity such that the transverse rho and axial-rho bubbles survive in the non relativistic limit. We thus conclude this discussion by summarizing the bare bubbles surviving in the non relativistic limit:
\begin{equation}
-\Pi^{NR}_{0aL}=-\Pi^{NR}_{0aT}=\Pi^{NR}_{0\rho R}	=\Pi_0,\qquad
\Pi^{NR}_{0\rho T}=-\left(\frac{1\,+\,\kappa^*_\rho}{2\, M^*}\right)^2\,\left|{\bf q}\right|^2\,\Pi_0,
\qquad \Pi^{NR}_{06}=-\frac{1\,+\,\kappa^*_\rho}{2\, M^*}\left|{\bf q}\right|\,\Pi_0 .
\end{equation}
In the non relativistic limit, the equations (\ref{VPI}) reduce to:
\begin{eqnarray}
(V_{aL}\,\Pi_{a L})^{NR}&=& \frac{V^{NR}_{L}\,\Pi_{0}}{1\,-\,V^{NR}_{L}\,\Pi_{0}}\nonumber\\
(V_{aT}\,\Pi_{a T}\,+\,V_{\rho}\,\Pi_{\rho })^{NR}&=& \frac{V^{NR}_{T}\,\Pi_{0}}{1\,-\,V^{NR}_{T}\,\Pi_{0}}\nonumber\\
(V_{\rho}\,\Pi_{\rho r})^{NR}_{}&=& \frac{V_\rho\,\Pi_{0}}{1\,-\,V_\rho\Pi_{0}}
\end{eqnarray}
where the non relativistic longitudinal and transverse spin-isospin interactions are given by:
\begin{eqnarray}
V^{NR}_{L}&=&\left(\frac{g_A}{2\,f_\pi}\right)^2\,\left({\bf q}^2\,D_\pi\,+\,g'\right)\,v^2(q)\nonumber\\
V^{NR}_{T}&=&\left(\frac{g_A}{2\,f_\pi}\right)^2\,\left({C_\rho\,\bf q}^2\,D_\rho\,+\,g'\right)\,v^2(q)
\quad\hbox{with}\quad C_\rho=\left(\frac{g_A}{2\,f_\pi}\right)^{-2}\,\left(\frac{1\,+\,\kappa^*_\rho}{2\, M^*}\right)^2\simeq 1.
\end{eqnarray}
Consequently, in this limit the correlation energy writes:
\begin{eqnarray}
E^{loop, NR}&=&- \frac{3 V}{2}\,\int_{-\infty}^{+\infty}{i\,d\omega\over 2\pi}\int{d{\bf q}\over (2\pi)^3}\,
\bigg[\ln\left(1\,-\,V^{NR}_{L}\,\Pi_{0}\right)\,+\,V^{NR}_{L}\,\Pi_{0}\bigg](\omega, {\bf q})\nonumber\\
& & - {3 V}\,\int_{-\infty}^{+\infty}{i\,d\omega\over 2\pi}\int{d{\bf q}\over (2\pi)^3}\,
\bigg[\ln\left(1\,-\,V^{NR}_{T}\,\Pi_{0}\right)\,+\,V^{NR}_{T}\,\Pi_{0}\bigg](\omega, {\bf q})\nonumber\\
& &- \frac{3 V}{2}\,\int_{-\infty}^{+\infty}{i\,d\omega\over 2\pi}\int{d{\bf q}\over (2\pi)^3}\,
\bigg[\ln\left(1\,-\,V_\rho\,\Pi_{0}\right)\,+\,V_\rho\,\Pi_{0}\bigg](\omega, {\bf q}).
\end{eqnarray}
The first two terms represent the longitudinal and transverse spin-isospin contributions as in ref. \cite{CE07}. The last term has nothing to do with spin-isospin physics but corresponds to the contribution of the time component of the vector interaction which is anyway very small ($-0.65\,MeV$ at normal nuclear matter density). As a byproduct we can estimate the contribution to the correlation energy of sigma and omega exchanges that we have not  considered explicitly in this paper. This omission can be justified with the following argument. In the non relativistic limit there is no difference between the scalar density and the vector density. The relevant interaction appearing in the RPA summation for the $\sigma + \omega$ channel is:
$$V_{\sigma + \omega}=\frac{g^{*2}_{S}}{q^2\,-\,m^{*2}_{\sigma}}\,-\,\frac{g^{2}_{\omega}}{q^2\,-\,m^{2}_{\omega}}.$$
Since the sigma, rho and omega have very similar masses around $800$ MeV and since the leading term to the correlation energy involve two meson exchanges lines, one can estimate
$$E^{loop}_{\sigma +\omega}\sim \frac{1}{3}\,\left(\frac{g^{2}_{\omega}\,-\,g^{*2}_{S}}{g^{2}_{\rho}}\right)^2\,E^{loop}_{\rho R}$$
where the $1/3$ is a trivial isospin factor. Taking $g_\omega=8.1$,
$g^{*}_{S}=6$, one obtains $E^{loop}_{\sigma +\omega}\sim -0.8$ MeV which is also very small and hence justifies the use of the mean-field approximation for the omega and sigma exchanges.

\begin{figure}
\begin{center}  
    \includegraphics[scale=0.5,angle=0]{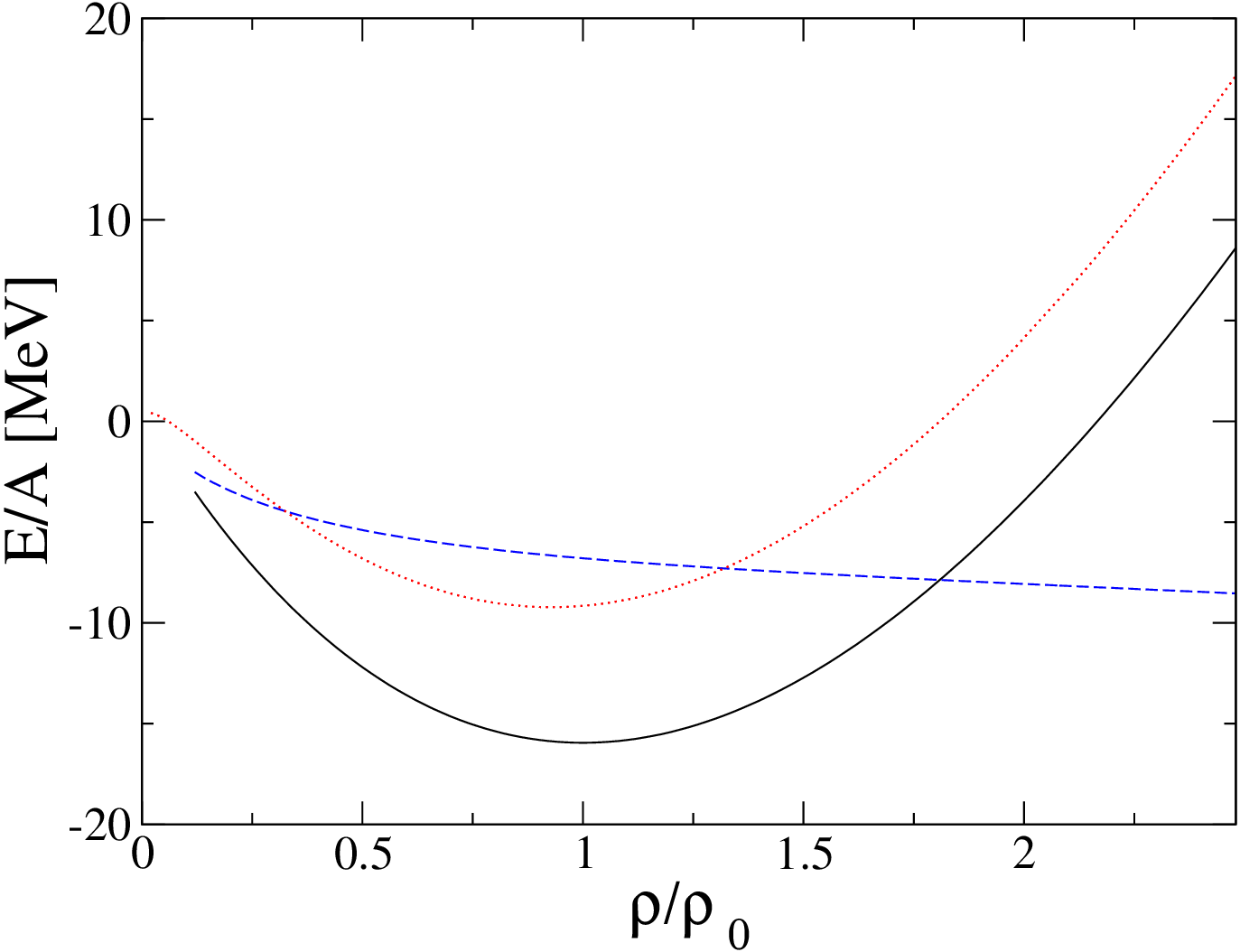}
 \end{center} 
  \caption{Binding energy of nuclear matter  with  $g'=0.7$. The values of the parameters and the
coordinates of the saturation point are given in Table \ref{tab:sat_rho}. The full line corresponds to the full result, the dotted line to the  mean-field (RHF) contribution and  the dashed line to the correlation energy.}
  \label{fig:sat_corr07}
\end{figure}
\begin{figure}[h]
  \begin{center}
    \includegraphics[scale=0.5,angle=0]{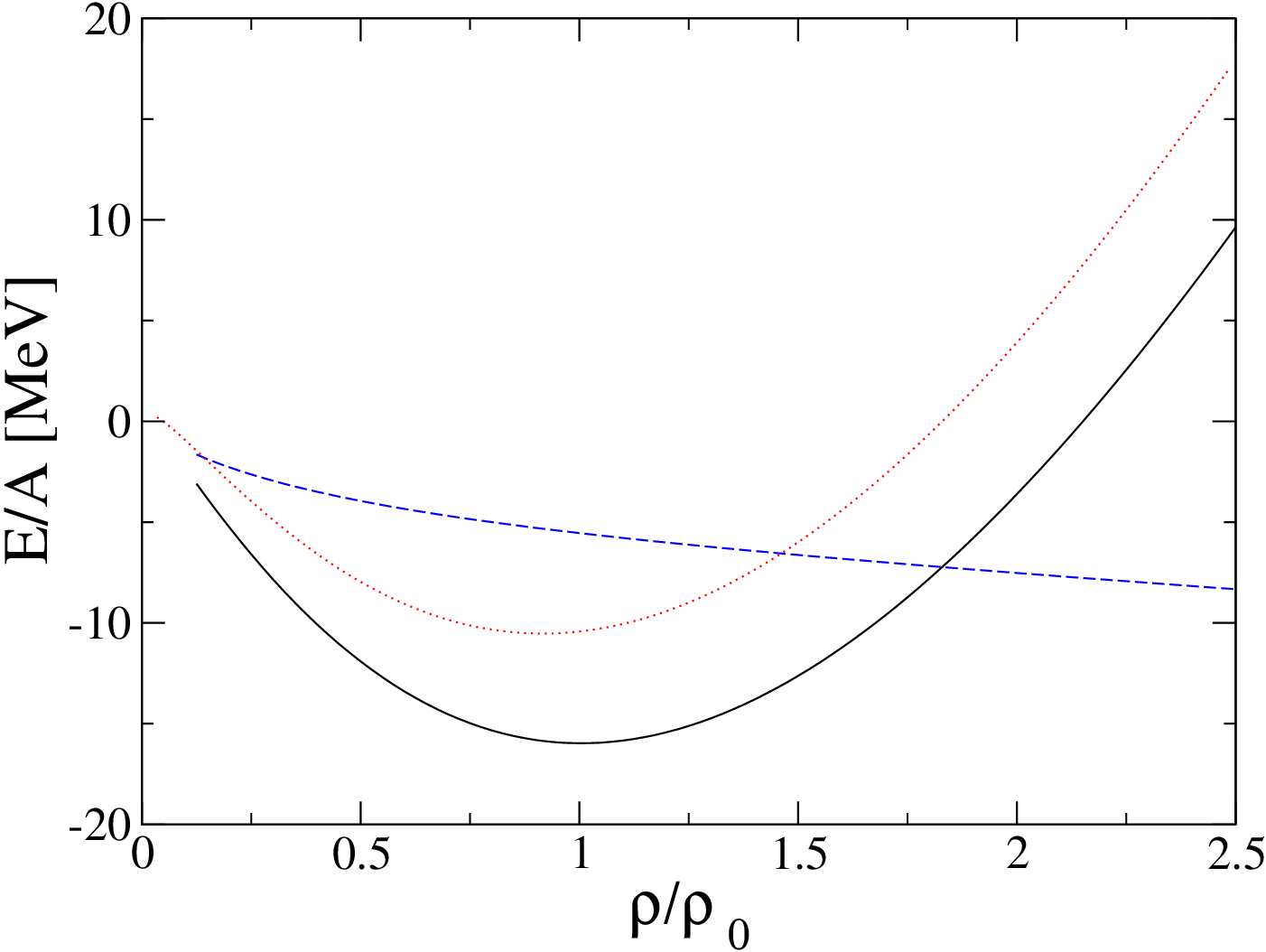}
  \end{center}
  \caption{ Same as figure 1 but for $g'=0.5$. The values of the parameters and the
coordinates of the saturation point are given in Table \ref{tab:sat_rho}.}
  \label{fig:sat_corr05}
\end{figure}

\bigskip
\paragraph{Numerical results and discussion.}

As explained before the spirit of the approach is to study nuclear matter properties with various parameters fixed as much as possible by hadron phenomenology and lattice data. We only allow a fine tuning for the $\omega NN$ coupling constant, $ g_\omega$, around the VDM/quark model value,
$3 \,g_\rho=7.95$ and for the nucleon scalar response parameter around the value estimated from the lattice, $C_{latt}\simeq 1.25$. The only nuclear physics input  is the Landau-Migdal parameter for which we first take the vastly used value $g'=0.7$ compatible with the most recent data analysis
\cite{IST06}. For $ g_\omega=7.6$ and $C=1.14 $ we obtain and excellent
reproduction of nuclear saturation properties as shown on fig.
\ref{fig:sat_corr07}. As compared with our previous pure Hartree-Fock work
\cite{MC08}  the $C$ parameter is reduced. Another  satisfactory result is the
value of the compressibility modulus $K=9 \rho^2\partial^2 \epsilon/\partial
\rho^2=250$ MeV,  smaller than our previous values and very close to the
accepted value around $240$ MeV. This can be understood by looking at figure
\ref{fig:sat_corr07} where the mean-field  and the correlation energy
contributions to the binding energy per nucleon are separately shown. The
important point is that the correlation energy displays a non linear behaviour
with density which helps saturation to occur: a smaller scalar response
parameter is needed hence reducing the incompressibility. It is also interesting
to examinate the various pieces contributing to the correlation energy. We see
on the last column of table \ref{tab:variousg} that, with the chosen value of
$g'$, there is an almost complete screening of the pion exchange and the
correlation energy is dominated by the transverse channel. At the saturation
point we find for the longitunal axial piece $E^{loop}_{aL}=-0.6$ MeV  whereas
the dominant transverse piece is $E^{loop}_{T}=-5.5$ MeV. We also see that the
axial R channel which is a pure relativistic correction is negligible:
$E^{loop}_{aR}=-0.04\, MeV$. Finally the relativistic correction to the rho
exchange energy which is not of spin-isospin nature is $E^{loop}_{\rho L}=-0.65$
MeV. 

\begin{table}
  \centering
  \begin{tabular}{|l|c|c|}
    \hline
    g' & 0.7 & 0.5 \\
    \hline
    $g_\omega$ & 7.6 & 7.3\\
    $C$ & 1.14 & 1.3\\
    $\rho/\rho_0$ & 1.00 & 1.00 \\
    \hline
    $E/A$ (MeV) & -15.97 & 15.87\\
    $K$ (MeV) & 250 & 270 \\
    \hline
  \end{tabular}  \caption{Values of the parameters and coordinates of the saturation point when
  the correlation energy is included on top of the RHF calculation.}
  \label{tab:sat_rho}
\end{table}

\begin{figure}
  \begin{tabular}{cc}
  \begin{minipage}{.50\linewidth}
    \includegraphics[scale=0.3,angle=0]{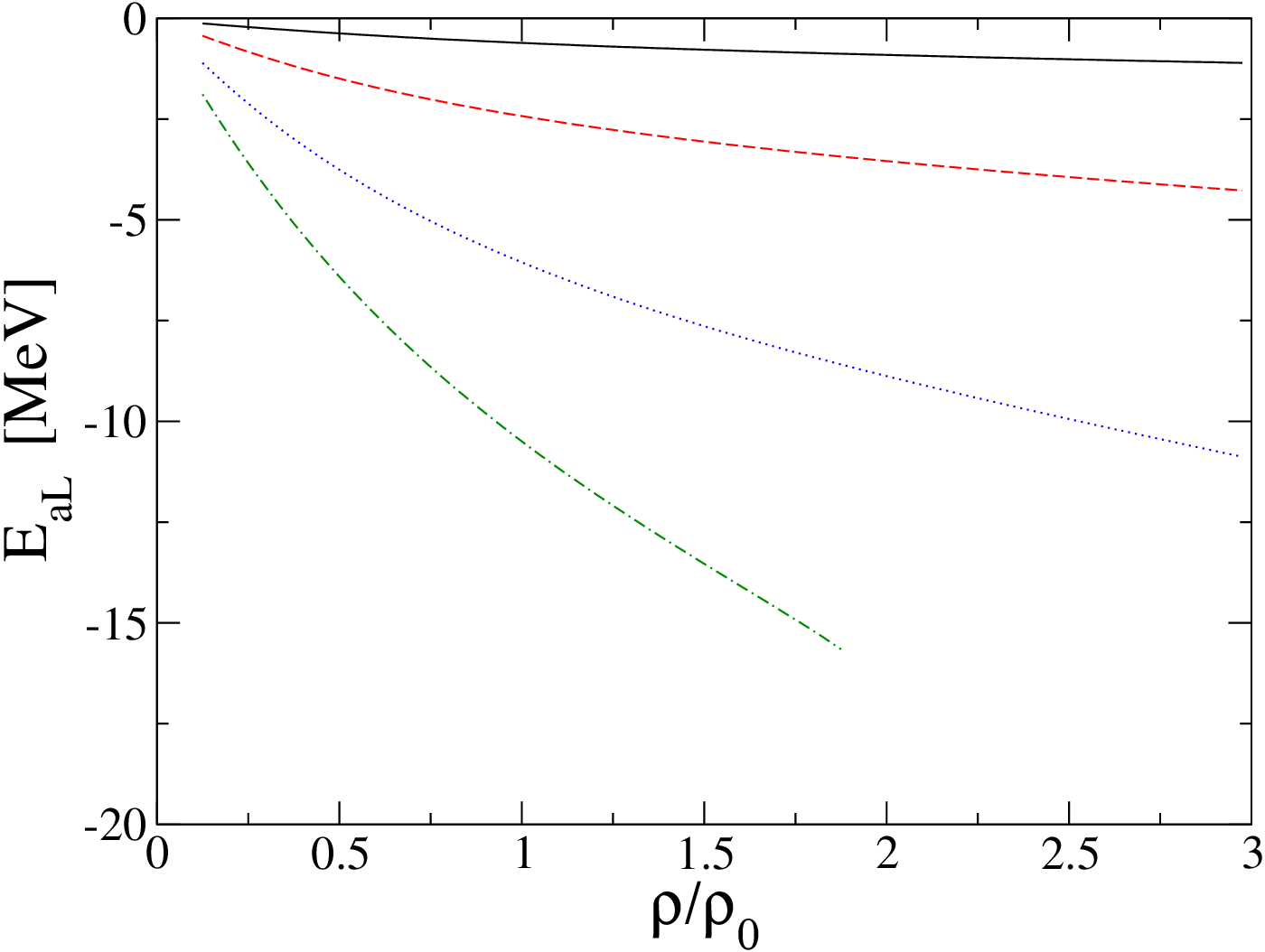}
  \end{minipage}
  &
  \begin{minipage}{.50\linewidth}
\includegraphics[scale=0.3,angle=0]{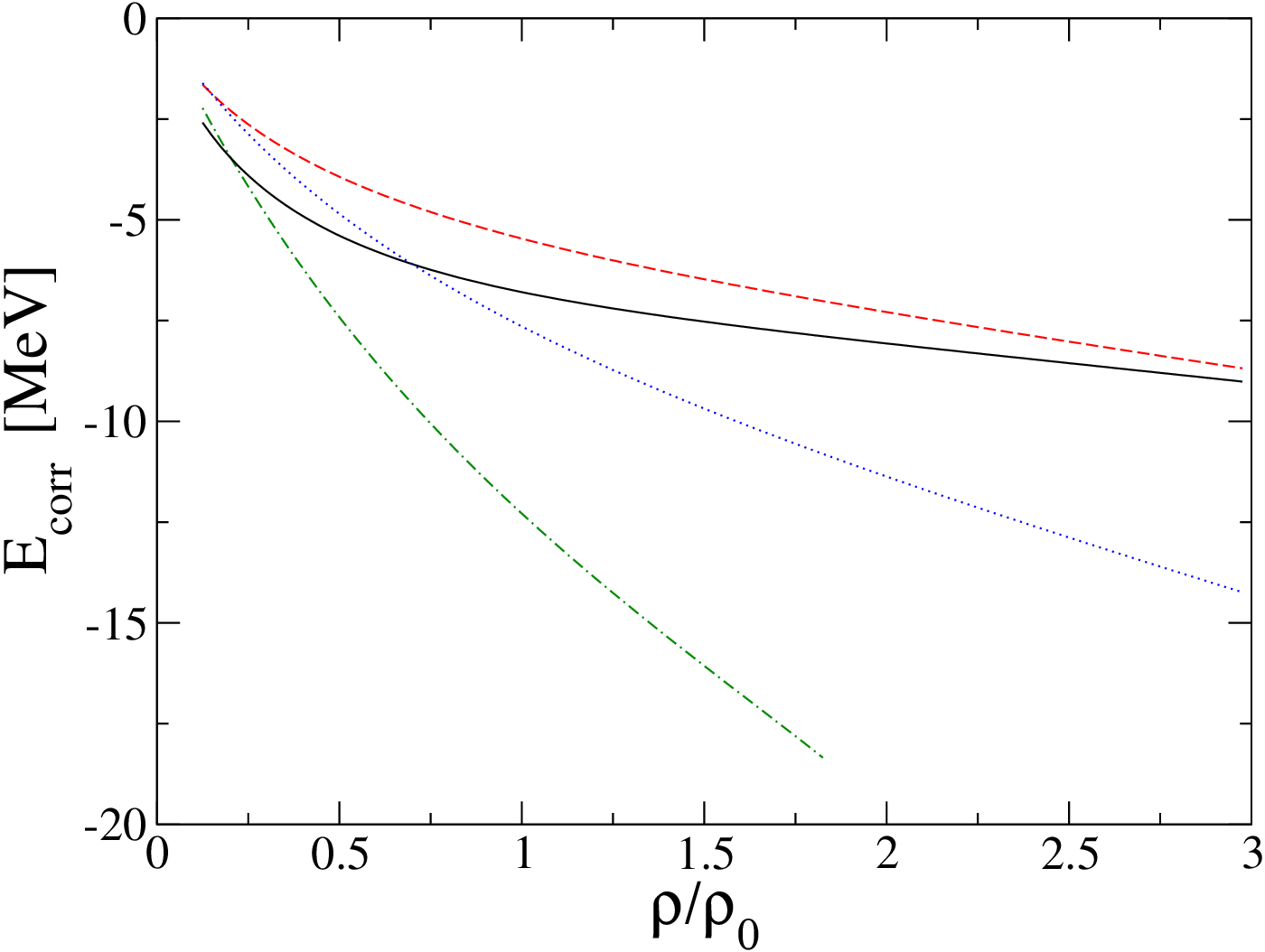}
   \end{minipage}
   \end{tabular}
   \caption{Longitudinal axial contribution to the correlation energy  (left panel) and total correlation energy (right panel) as a function of density for various values of $g'$. Solid curves: $g'=0.7$; dashed curves: $g'=0.5$: dotted curves: $g'=0.3$; dot-dashed curves: $g'=0.15$.}
   \label{DECOMP}
   \end{figure}

It is also interesting  to discuss the influence of some inputs of the
calculation although they are constrained by accepted phenomenology. If the $g'$
parameter is decreased to the value $g'=0.5$ the saturation properties can be
also reproduced (see fig
\ref{fig:sat_corr05}) but the compressibilty modulus is larger: $K=270$ MeV
(see Table \ref{tab:sat_rho}).  In that case the  screening of the pion exchange
is less pronounced $E^{loop}_{aL}=-2.4$ MeV but the transverse piece is
reduced $E^{loop}_{T}=-2.1$ MeV. Actually 
this transverse contribution exhibits a minimum around $g'=0.2$ which
corresponds to the strongest compensation between attractive rho exchange and
the contact interaction. This can be seen on Table \ref{tab:variousg} where the
various contributions to the correlation energie are given at the normal
nuclear matter density for various values of $g'$. An important point is that
decreasing  $g'$ brings nuclear matter close to pion condensation \cite{MIG78}.
The onset of pion condensation corresponds to a pole at zero energy in the pion
propagator or in the axial longitudinal polarization propagator at some critical
momentum $q_c$, namely $(1\,-\,V_{aL}\,\Pi_{0a L})(\omega=0, q_c)=0$. This is
illustrated on the left panel of fig. \ref{DECOMP} where $E^{loop}_{aL}$ is
displayed for various values of $g'$. For the case $g'=0.15$ the calculation
ends at a density $\rho=1.8\,\rho_0$ which is the critical density for pion
condensation. It turns out that the relativistic calculation disfavors pion
condensation as compared with the non relativistic one. For instance in the non
relativistic calculation pion condensation occurs at $\rho=2.2\,\rho_0$ for
$g'=0.3$ whereas it is beyond $3 \,\rho_0$ in the relativistic case. Close to
pion condensation the longitudinal correlation energy strongly increases which
is reminiscent of a critical opalescence phenomena \cite{DE78}.

\noindent
   \begin{table}[h]
\begin{center}
\begin{tabular}{|l|c|c|c|c|c|c|c|c|}
\hline
& $g'=0$ & $g'=0.1$ & $g'=0.2$ & $g'=0.3$ & $g'=0.4$ &$g'=0.5$ &$g'=0.6$ &$g'=0.7$ \\
\hline 
$E^{loop}_{aL}$ & x & 12.5 & -8.8 & -6.0 & -4.0 & -2.4 & -1.3 & -0.6  \\
\hline
$E^{loop}_{T}$ & -2.6 & -1.5 & -0.9& -0.9 & -1.4 & -2.1 & -3.7 & -5.5 \\
\hline 
$E^{loop}_{aL}$  & -0.00 & -0.00 & -0.00 & -0.01 & -0.01 & -0.02& -0.03 & -0.04 \\
\hline
$E^{loop}_{\rho R}$ & -0.65  & -0.65 & -0.65 & -0.65 & -0.65 & -0.65 & -0.65 & -0.65 \\
\hline
\end{tabular}
\end{center}
\caption{The various contributions (MeV) to the correlation energy at normal nuclear matter density for various values of  $g'$ with $C=1.14$ and $g_\omega=7.6$. The cross for the longitudinal contribution for the $g'=0$ case means that pion condensation occurs.}
\label{tab:variousg}
\end{table}

The magnitude  of the correlation energy is  controlled by the cutoffs entering
the form factors. For the $\pi N$ form factor, it   is fixed by the pion cloud
contribution to the pion-nucleon sigma term. However there is no real reason to
keep for the rho exchange contribution the same form factor. Therefore  we have
changed  the cutoff entering the $\rho N$ form factor from
$\Lambda_\pi=980$ MeV to $\Lambda_\rho=1500$ MeV. It turns out that the transverse contribution is not  much increased and the saturation curve
can be obtained with only a tiny change of the $C$ and $g_\omega$ parameters, the main effect being the increase by a factor three of the relativistic correction, $E^{loop}_{\rho R}$ to the rho meson exchange. This is illustrated in Table \ref{tab:newrhocutoff}.  

However even if this form factor may seem to be quite  soft we also have to keep in mind that a more careful treatment of short range correlation generates a momentum dependence of the $g'$ parameter together with a tensor $h'$ parameter in such a way that the residual interactions in both channels vanish at high momentum (Beg-Agassi-Gal theorem) hence accelarating the convergence of the momentum integration \cite{CDE85}. Another point is the neglect of  delta-hole bubbles which contribute significantly to the correlation energy according to our previous non relativistic calculation \cite{CE07}. However we found in \cite{CE07} that it gives a contribution 
almost exactly linear in density which is phenomenologically undistinguishable from omega exchange. In practice a very small increase of $g_\omega$ is expected to simulate this delta-hole bubble contribution.  

\begin{table}
  \centering
  \begin{tabular}{|l|c|c|c|}
    \hline
    $\Lambda_\rho\,[MeV]$  & 980 & 1500 & 1500 \\
    \hline
    $g_\omega$ & 7.6 & 7.6 & 7.8\\
    $C$ & 1.14 & 1.14 & 1.15\\
    \hline
    $K\,[MeV]$ & 250 &  & 270 \\
    $E^{loop}_{T}\,[MeV]$ &-6.8 & -9.4 & -9.4\\
    $E^{loop}_{\rho R}\,[MeV]$ & -0.6 & -1.6 & -1.6 \\
    \hline
  \end{tabular}
  \caption{Dependence of the correlation energy with the cutoff entering the $\rho N$ form factor. The first column corresponds to the original calculation with $g'=0.7$ and the second column gives the modification of the correlation energy when changing the cutoff. The last column 
  shows the new parameters when the saturation point is readjusted.}
  \label{tab:newrhocutoff}
\end{table}

\section{Conclusion}
In this work we have developped a fully relativistic RPA framework for the calculation of the correlation energy.
We have included the effect of pion loops, as well as the rho ones, on top of a RHF calculation. Our description 
of saturation properties is almost parameter free. The nuclear binding is ensured by a background chiral invariant scalar field associated with
the radial fluctuations of the quark condensate. In order to reach saturation we have incorporated
the response of the embedded nucleon to the background scalar field. It generates three-body repulsive forces  which allows saturation to occur. All the parameters entering the Hartree-Fock description, including the scalar nucleon response, have been fixed or constrained by hadron or QCD phenomenology. 

The Landau-Migdal parameter $g'$ entering the correlation energy is taken from the phenomenology of spin-isospin physics. We have found that  the magnitude of this  correlation energy is  moderate, of the order $-10$ MeV per nucleon,   due to the  strong screening of the pion exchange by short range correlations. Eventhough  this contribution is not very large its non linear  behaviour with density improves the result of the RHF mean-field calculation for the description of bulk properties of nuclear matter. This confirms our previous conclusion obtained in a non relativistic evaluation of the correlation effects \cite{CE07}.  This result seems to be  robust in the sense that this conclusion is not altered if we vary  the $\rho N$ form factor or the $g'$ parameter around $g'=0.7$,  the standard value.  
 \vskip 1 true cm
{\bf Acknowledgments.} We would like to thank M. Ericson for constant interest in this work and critical reading of the manuscript. We have also benefited from discussions with H. Hansen and M. Martini.  
   
\vfill\eject

\appendix 
{\bf \Large{APPENDIX}: Explicit form of the  bare polarization propagators after Wick rotation}

\vskip 1 true cm

In this appendix we give the expressions for the various mean-field polarization propagators once the integration over the angle between the transferred moment ${\bf q}$ and  the nucleon momentum ${\bf p}$  has been performed.
 They are given after the Wick
rotation (eq. \ref{eqn:WICKrot}) which is used to perform the practical calculation of the correlation energy.
We use the following tool functions inspired from \cite{LP03}~:
\begin{eqnarray}
  \ln_1(z,|{\bf q}|) &=& \ln\left( \frac{(q^2-2|{\bf p}||{\bf
  q})^2+4z^2E_p^2}{(q^2+2|{\bf p}||{\bf   q}|)^2+4z^2E_p^2}
  \right) \nonumber \\
  z\ln_2(z,|{\bf q}|) &=&  -2 \,z\, atan\left( \frac{8|{\bf p}||{\bf
  q}|zE_p}{(q^2)^2-4{\bf p}^2{\bf   q}^2+4z^2E_p^2} \right). \nonumber
  \label{}
\end{eqnarray}
The expressions of the various propagators are~:
\begin{eqnarray}
  \Pi_{0 a T} &=&  -\frac{1}{8\pi^2}\frac{q^2}{|{\bf
  q}|^3}\int\frac{p\,dp}{E_p}\left\{ \left( 4E_p^4-z^2+|{\bf
  q}|^2-4\frac{|{\bf q}|^2}{q^2}M^{*2} \right)\ln_1(z,|{\bf q}|)+8|{\bf p}||{\bf
  q}|\frac{-z^2+|{\bf q}|^2}{q^2}+4E_p z\ln_2(z,|{\bf q}|) \right\}
  \nonumber\\
  \Pi_{0 a R} &=& \frac{1}{4\pi^2}\frac{1}{|{\bf q}|^3}
  \int\frac{p\,dp}{E_p} \left\{ \left(
  -4M^{*2}z^2+4|{\bf p}|^2q^2+(q^2)^2
  \right)\ln_1(z,|{\bf q}|) +8|{\bf p}||{\bf  q}|q^2 +4E_p z
  \ln_2(z,|{\bf q}|)\right\}
  \nonumber\\
  \Pi_{0 a L} &=& \frac{1}{4\pi^2}\frac{1}{|{\bf q}|} \int\frac{p\,dp}{E_p}
  4M^{*2}\ln_1(z,|{\bf q}|)
  \nonumber\\
  \Pi_{0 \rho T} &=& \frac{1}{4\pi^2}\frac{q^2}{|{\bf q}|^3}
  \int\frac{p\,dp}{E_p} \bigg[ -\frac{1}{2}\left\{ \left(
  4E_p^2-z^2+|{\bf q}|^2+4\frac{|{\bf q}|^2}{q^2}M^{*2}
  \right)\ln_1(z,|{\bf q}|)+8|{\bf p}||{\bf  q}|\frac{-z^2+|{\bf q}|^2}{q^2}+ 4E_p z
  \ln_2(z,|{\bf q}|)\right\} \nonumber\\
  & &\, -\kappa_\rho^*|{\bf q}|^2\ln_1(z,|{\bf q}|)
  +\frac{1}{2}\left( \frac{\kappa_\rho}{2M_N} \right)\left\{ \left(
  -4M^{*2}|{\bf q}|^2+4E_p^2q^2+(q^2)^2
  \right)\ln_1(z,|{\bf q}|) +8|{\bf p}||{\bf  q}|q^2 +4E_p z
    \ln_2(z,|{\bf q}|)\right\} \bigg]
  \nonumber\\
  \Pi_{0 \rho R} &=& \frac{1}{4\pi^2}\frac{q^2}{|{\bf q}|^3}
  \int\frac{p\,dp}{E_p} \bigg[ \left\{ \left( 4E_p^2+q^2
  \right)\ln_1(z,|{\bf q}|) 8|{\bf p}||{\bf  q}|+4E_p z
  \ln_2(z,|{\bf q}|)\right\}-2\kappa_\rho^*|{\bf q}|^2\ln_1(z,|{\bf q}|) \nonumber\\
  & & \, +\left(\frac{\kappa_\rho}{2M_N} \right) \left\{ \left( z^2|{\bf q}|^2 +4{\bf p}^2{\bf
  q}^2+4E_p^2 z^2\right)\ln_1(z,|{\bf q}|) +8|{\bf p}||{\bf  q}|z^2
  -4q^2E_p z\ln_2(z,|{\bf q}|) \right\}\bigg]
  \nonumber\\
  \Pi_{06} &=& \frac{\kappa_\rho^*}{4\pi^2}\frac{q^2}{|{\bf q}|^2}
  \int\frac{p\,dp}{E_p} \left\{ 2E_p\ln_1(z,|{\bf q}|)
  +z\ln_2(z,|{\bf q}|) \right\}.
  \label{APP}
\end{eqnarray}


\begin{thebibliography}{99}
\bibitem{CE05} G. Chanfray and M. Ericson, EPJA 25 (2005) 151.
\bibitem{CE07} G. Chanfray and M. Ericson, Phys. Rev C75 (2007) 015206.
\bibitem{MC08} \'E. Massot and G. Chanfray, Phys. Rev. C78 (2008) 015204.
\bibitem{CHS01} L.S. Celenza, Huangsheng Wang and C.M. Shakin, Phys. Rev. C763 (2001) 025209.
\bibitem{BT01} W. Bentz and A.W. Thomas, Nucl. Phys. A696 (2001) 138.
\bibitem{CEG02} G. Chanfray, M. Ericson and P.A.M. Guichon, Phys. Rev C63 (2001)055202.
\bibitem{SW86} B.D. Serot, J.D. Walecka, Adv. Nucl. Phys. 16 (1986) 1; Int. J. Mod. Phys. E16 (1997) 15.
\bibitem{KM74} A.K. Kerman and L.D. Miller in ``Second High Energy Heavy Ion Summer Study'',
LBL-3675, 1974. 
\bibitem{EC07}M. Ericson and G. Chanfray, EPJA 34 (2007) 215.
\bibitem{G88} P.A.M. Guichon, Phys. Lett. B200 (1988) 235.
\bibitem{LTY04} D.B. Leinweber, A.W. Thomas and R.D. Young, Phys. Rev. Lett 92 (2004) 242002.
\bibitem{N01} M. Nakano {\it et al}, Int. J. Mod. Phys. E10 (200) 459.
\bibitem{L03} M.F.M. Lutz, Phys. Lett. B552 (2003) 159.
\bibitem{JTC92} I. Jameson, A.W. Thomas and G. Chanfray, J. Phys. G18, (1992) L159 .
\bibitem{BM92} M.C. Birse and J.E. McGovern, Phys. Lett. B292  (1992) 242. 
\bibitem{TGLY04} A. W. Thomas, P. A. M. Guichon, D. B.  Leinweber and R. D. Young
Progr. Theor. Phys. Suppl. 156, 124 (2004); nucl-th/0411014. 
\bibitem{CPS92} L.S. Celenza, A. Pantziris and C.M. Shakin, Phys. Rev. C45 (19992) 205.
\bibitem{KFW02} N. Kaiser, S. Fritsch and W. Weise, Nucl. Phys. A697 (2002) 82.
\bibitem{CL99} J.C. Caillon and J. Labarsouque, Phys. Rev. C59 (1999) 1090.
\bibitem{IST06} M. Ichimura, H. Sakai and T. Wakasa, Progr. Part. Nucl. Phys. 56 (2006) 446.
\bibitem{MIG78} A.B. Migdal, Rev. Mod. Phys. 50 (1978) 107.
\bibitem{DE78} J. Delorme and M. Ericson, Phys. Lett. B76 (1978) 182.
\bibitem{CDE85}G. Chanfray, J. Delorme and M. Ericson, Phys. Rev. C31 (1985) 1582.
\bibitem{LP03}L.P. Leinson and A. P\'erez, Ar.XiV nucl-th/0307025.


\end{thebibliography}
\end{document}